\documentclass[12pt]{article}
\usepackage{amsmath,amssymb,epsfig,latexsym,txfonts}

\begin{document}

\begin{titlepage}

\begin{flushright}
CLNS~05/1948\\
FERMILAB-PUB-05-537-T\\[0.2cm]
December 15, 2005
\end{flushright}

\vspace{0.7cm}
\begin{center}
\Large\bf
\boldmath
Toward a NNLO calculation of the $\bar B\to X_s\gamma$ decay rate with a cut
on photon energy:\\
I. Two-loop result for the soft function
\unboldmath
\end{center}

\vspace{0.8cm}
\begin{center}
{\sc Thomas Becher$^a$ and Matthias Neubert$^{b,c}$}\\
\vspace{0.4cm}
{\sl $^a$\,Fermi National Accelerator Laboratory\\
P.O. Box 500, Batavia, IL 60510, U.S.A.\\[0.3cm]
$^b$\,Institute for High-Energy Phenomenology\\
Newman Laboratory for Elementary-Particle Physics, Cornell University\\
Ithaca, NY 14853, U.S.A.\\[0.3cm]
$^c$\,Institut f\"ur Theoretische Physik, Universit\"at Heidelberg\\
Philosophenweg 16, D--69120 Heidelberg, Germany}
\end{center}

\vspace{1.0cm}
\begin{abstract}
\vspace{0.2cm}
\noindent 
A theoretical analysis of the partial inclusive $\bar B\to X_s\gamma$ decay 
rate with a cut $E_\gamma\ge E_0$ on photon energy must deal with 
short-distance contributions associated with three different mass scales: the 
hard scale $m_b$, an intermediate scale $\sqrt{m_b\Delta}$, and a soft scale 
$\Delta$, where $\Delta=m_b-2E_0\approx 1$\,GeV for $E_0\approx 1.8$\,GeV. 
The cut-dependent effects are described in terms of two perturbative objects 
called the jet function and the soft function, which for a 
next-to-next-to-leading order analysis of the decay rate are required with
two-loop accuracy. The two-loop calculation of the soft function is presented
here, while that of the jet function will be described in a subsequent paper. 
As a by-product, we rederive the two-loop anomalous-dimension kernel of the 
$B$-meson shape function.
\end{abstract}
\vfil

\end{titlepage}

\section{Introduction}
Weak-decay processes involving flavor-changing neutral currents are sensitive 
to the effects of new physics, because the decay amplitudes are 
loop-suppressed in the Standard Model. In this context, the decay 
$\bar B\to X_s\gamma$ plays an especially prominent role, since its rate is 
being measured increasingly well by the $B$-factories. The current 
experimental precision already matches the theoretical accuracy of the 
next-to-leading logarithmic prediction. This has triggered an effort to push 
the precision of the theoretical calculation of the decay rate in the Standard 
Model to the next level of accuracy. Due to the presence of several different 
scales in the decay process, this calculation involves a number of different 
elements. 

Several of the steps required to achieve next-to-next-to-leading logarithmic 
(NNLO) accuracy have already been taken. The matching of the Standard Model 
onto an effective weak Hamiltonian has been completed by performing a 
three-loop matching calculation onto the electro- and chromo-magnetic dipole 
operators \cite{Misiak:2004ew}. The effective weak Hamiltonian allows one to 
resum large perturbative logarithms of the form $\alpha_s\ln(M_W/m_b)$. To 
this end, the three-loop anomalous-dimension matrices for the four-quark 
operators \cite{Gorbahn:2004my} and for the mixing of the dipole operators 
among each other \cite{Gorbahn:2005sa} have been calculated. The evaluation of 
the four-loop anomalous dimension of the mixing of the current-current 
operators into the dipole operators is in progress and is now the only missing 
element to obtain the Wilson coefficients in the effective weak Hamiltonian 
with NNLO precision. The most difficult part of the calculation is the 
evaluation of the matrix elements of the corresponding operators, in 
particular the ones involving penguin contractions of current-current 
operators with charm-quark loops. So far, the complete two-loop matrix element 
is known only for the electro-magnetic dipole operator $Q_{7\gamma}$ 
\cite{Blokland:2005uk}. For other operators, only the parts proportional to 
$\beta_0\alpha_s^2$ (more precisely, the terms proportional to 
$n_f\alpha_s^2$) have been obtained \cite{Bieri:2003ue}. For the case of 
$Q_{7\gamma}$, not only the total decay rate but also the photon-energy 
spectrum has been evaluated at two-loop order \cite{Melnikov:2005bx}. For the 
remaining operators in the effective weak Hamiltonian, the photon spectrum is 
known to order $\beta_0\alpha_s^2$ \cite{Ligeti:1999ea}.

An important motivation for undertaking a NNLO (i.e., order $\alpha_s^2$) 
evaluation of the $\bar B\to X_s\gamma$ decay rate is the fact that such a 
calculation would reduce the strong dependence on the renormalization scheme 
adopted for the charm-quark mass, which is observed at NLO 
\cite{Gambino:2001ew, Asatrian:2005pm}. However, the charm mass is not the 
only low scale in the decay process. Another set of enhanced corrections 
arises because it is experimentally necessary to put a cut $E_\gamma>E_0$ on 
the photon energy (defined in the $B$-meson rest frame). The relevant scale is 
$\Delta=m_b-2E_0$. For $E_0\approx 1.8$\,GeV, the currently lowest value of 
the cut achieved by the Belle experiment \cite{Koppenburg:2004fz}, the scale 
$\Delta\approx 1$\,GeV is barely in the perturbative domain. For even higher 
values of $E_0$, the effects associated with the scale $\Delta$ cannot be 
calculated reliably in perturbation theory, in which case they are relegated 
into a nonperturbative shape-function \cite{Neubert:1993ch,Kagan:1998ym}.
 
As long as the cut energy is chosen sufficiently low, such that 
$m_b\gg\Delta\gg\Lambda_{\rm QCD}$, the partial inclusive 
$\bar B\to X_s\gamma$ decay rate can be calculated perturbatively using a 
multi-scale operator-product expansion. It consists of a simultaneous 
expansion in powers of $\Delta/m_b$ and $\Lambda_{\rm QCD}/\Delta$, combined 
with a systematic resummation of logarithms of ratios of the hard, 
intermediate, and soft scales \cite{Neubert:2004dd} (see also 
\cite{Bauer:2001yt}). At leading power in $\Delta/m_b$ and next-to-leading 
order in the expansion in powers of $\Lambda_{\rm QCD}/\Delta$, it is possible 
to derive an exact expression for the partial decay rate $\Gamma(\Delta)$, 
valid to all orders in perturbation theory, in which the dependence on the 
variable $\Delta$ enters in a transparent way. The result is 
\cite{Neubert:2005nt}
\begin{eqnarray}\label{GOPE}
   &&\hspace{-0.6cm}\Gamma(\Delta)
    = \frac{G_F^2\alpha}{32\pi^4}\,|V_{tb} V_{ts}^*|^2\,m_b^3\,
    \overline{m}_b^2(\mu_h)\,|H_\gamma(\mu_h)|^2\,
    U_1(\mu_h,\mu_i)\,U_2(\mu_i,\mu_0)
    \left( \frac{\Delta}{\mu_0} \right)^\eta\\
   &\times& \left\{ \widetilde j\Big( \ln\frac{m_b\Delta}{\mu_i^2}
    + \partial_\eta,\mu_i \Big)\,
    \widetilde s\Big( \ln\frac{\Delta}{\mu_0} + \partial_\eta,\mu_0 \Big)\, 
    \frac{e^{-\gamma_E\eta}}{\Gamma(1+\eta)} \left[
    1 - \frac{\eta(1-\eta)}{6}\,\frac{\mu_\pi^2}{\Delta^2} + \dots \right]
    + {\cal O}\left( \frac{\Delta}{m_b} \right) \right\} . \nonumber
\end{eqnarray}
Here $m_b$ is the $b$-quark pole mass, and $\overline{m}_b(\mu)$ denotes the 
running mass defined in the $\overline{\rm MS}$ scheme. The only hadronic 
parameter entering at this order is the quantity $\mu_\pi^2$ related to the 
$b$-quark kinetic energy inside the $B$ meson. The ellipses represent 
subleading corrections of order $(\Lambda_{\rm QCD}/\Delta)^3$, which are 
unknown. The pole mass and $\mu_\pi^2$ must be eliminated in terms of related 
parameters defined in a physical subtraction scheme, such as the 
shape-function scheme \cite{Bosch:2004th,Neubert:2004sp}. The scales 
$\mu_h\sim m_b$, $\mu_i\sim\sqrt{m_b\Delta}$, and $\mu_0\sim\Delta$ are hard, 
intermediate, and soft matching scales. The hard function $H_\gamma$, the jet 
function $\widetilde j$, and the soft function $\widetilde s$ encode the 
contributions to the rate associated with these scales. Note that all 
information about the short-distance quantum fluctuations associated with the 
weak-interaction vertices in the effective weak Hamiltonian are contained in 
$H_\gamma$. Logarithms of ratios of the various scales are resummed into the 
evolution functions $U_1$ (evolution from the hard to the intermediate scale) 
and $U_2$ (evolution from the intermediate to the soft scale), as well as into 
the quantity 
\begin{equation}\label{etadef}
   \eta = 2\int_{\mu_0}^{\mu_i} \frac{d\mu}{\mu}\,
   \Gamma_{\rm cusp}[\alpha_s(\mu)] \,,
\end{equation}
which is given in terms of an integral over the universal cusp anomalous 
dimension of Wilson loops with light-like segments \cite{Korchemsky:1987wg}.
The result (\ref{GOPE}) is formally independent of the choices of the matching 
scales. In practice, a residual scale dependence remains because one is forced 
to truncate the perturbative expansions of the various objects in the formula 
for the decay rate. Reducing the scale uncertainty associated with the lowest
short-distance scale, $\Delta\approx 1$\,GeV, is the goal of the present work.

The soft function $\widetilde s$ in (\ref{GOPE}) is related to the original 
$B$-meson shape function $S(\omega,\mu)$ \cite{Neubert:1993ch} through a 
series of steps. Starting from a perturbative calculation of the shape 
function in the parton model with on-shell $b$-quark states, we first define
\begin{equation}\label{jdefsdef}
   s\Big( \ln\frac{\Omega}{\mu},\mu \Big)
   \equiv \int_0^{\Omega} d\omega\,S_{\rm parton}(\omega,\mu) \,.
\end{equation}
For $\Omega\gg\Lambda_{\rm QCD}$, this parton-model expression gives the 
leading term in a systematic operator-product expansion of the integral over 
the true shape function \cite{Neubert:2005nt}. The first power correction is 
linked to the leading term by reparameterization invariance 
\cite{Luke:1992cs,Falk:1992fm} and gives rise to the term proportional to 
$\mu_\pi^2/\Delta^2$ in (\ref{GOPE}). While the perturbative expression for 
$S_{\rm parton}$ involves singular distributions \cite{Bosch:2004th}, the 
function $s$ has a double-logarithmic expansion of the form
\begin{eqnarray}\label{sexp}
   s(L,\mu) &=& 1 + \sum_{n=1}^\infty
    \left( \frac{\alpha_s(\mu)}{4\pi} \right)^n
    \left( c_0^{(n)} + c_1^{(n)} L + \dots + c_{2n-1}^{(n)} L^{2n-1}
    + c_{2n}^{(n)} L^{2n} \right) .
\end{eqnarray}
The function $\widetilde s$ is then obtained by the replacement rule 
\cite{Neubert:2005nt}
\begin{equation}\label{replace}
   \widetilde s(L,\mu)\equiv s(L,\mu) \Big|_{L^n\to I_n(L)} \,, 
\end{equation}
where $I_n(x)$ are $n$-th order polynomials defined as
\begin{equation}
   I_n(x) = \frac{d^n}{d\epsilon^n}\,
   \exp\left[ \epsilon x + \sum_{k=2}^\infty\,
   \frac{(-1)^k}{k}\,\epsilon^k \zeta_k \right] \Bigg|_{\epsilon=0} .
\end{equation}
By solving the renormalization-group equation for the soft function order by 
order in perturbation theory, the coefficients $c_{k\ne 0}^{(n)}$ of the 
logarithmic terms in (\ref{sexp}) can be obtained from the expansion 
coefficients of the shape-function anomalous dimension and the 
$\beta$-function, together with the coefficients $c_0^{(n)}$ coefficients 
arising in lower orders \cite{Neubert:2005nt}. The two-loop calculation 
performed in the present paper gives the constant $c_0^{(2)}$ and provides a 
check on the two-loop anomalous dimension of the shape function. We also note 
that from our result for $s(L,\mu)$ one can derive the two-loop expression for 
$S_{\rm parton}(\omega,\mu)$ in terms of so-called star distributions \cite{Bosch:2004th}.

In the next section, we discuss how to perform the two-loop calculation of the 
soft function $s$ in an efficient way. The calculation is simplified by 
representing the $\delta$-function operator appearing in the shape-function as 
the imaginary part of a light-cone propagator. In this way, we avoid having to 
deal with distribution-valued loop integrals and instead map the calculation 
to the evaluation of on-shell two-loop integrals with heavy-quark and 
light-cone propagators. Using integration-by-parts relations among these loop 
integrals, the entire calculation is reduced to the evaluation of four master 
integrals. After presenting the result for the bare soft function, we discuss 
its renormalization in Section \ref{sec:renormalization}. The relevant 
anomalous dimension depends both implicitly (through the coupling constant) 
and explicitly (through a star distribution) on the renormalization scale. 
This explicit dependence gives rise to Sudakov logarithms in the soft 
function. We conclude after presenting our final expression for the 
renormalized soft function in Section~\ref{sec:concl}.

\section{Two-loop calculation of the soft function}

The definition of the soft function $s$ in (\ref{jdefsdef}) implies that
\begin{equation}
   s\Big( \ln\frac{\Omega}{\mu},\mu \Big)
   \equiv \int_0^{\Omega} d\omega\,
   \langle b_v|\,\bar h_v\,\delta(\omega+in\cdot D)\,h_v\,|b_v\rangle \,,
\end{equation}
where $h_v$ are effective heavy-quark fields in heavy-quark effective theory 
\cite{Neubert:1993mb}, $b_v$ are on-shell $b$-quark states with velocity $v$, 
and $n$ is a light-like 4-vector satisfying $n\cdot v=1$ (note that $v^2=1$ 
and $n^2=0$). The normalization of states is such that 
$\langle b_v|\,\bar h_v\,h_v\,|b_v\rangle=1$.

Working with the above representation of the soft function is difficult due to 
the presence of the $\delta$-function differential operator, the Feynman rules 
for which involve $\delta$ functions and their derivatives. This complication 
can be avoided by writing the $\delta$-function operator as the discontinuity
of a light-cone propagator in the background of the gluon field. This allows 
us to represent the soft function as a contour integral in the complex 
$\omega$ plane:
\begin{equation}\label{eq:softdisp}
   s\Big( \ln\frac{\Omega}{\mu},\mu \Big)
   = \frac{1}{2\pi i} \ointctrclockwise\limits_{|\omega|=\Omega} d\omega\,
   \langle b_v|\,\bar h_v\,\frac{1}{\omega+in\cdot D+i0}\,h_v\,|b_v\rangle
   \,. \vspace{-0.1cm}
\end{equation}
The Feynman rules for the gauge-covariant propagator involve light-cone 
propagators of the type $(\omega+n\cdot p)^{-1}$, which are straightforward to
deal with using dimensional regularization and standard loop techniques. 
Dimensional analysis implies that an $n$-loop contribution to the matrix 
element in the integrand of the contour integral is proportional to 
$(-\omega)^{-1-2n\epsilon}$, where $d=4-2\epsilon$ is the dimension of 
space-time. The relevant contour integration yields
\begin{equation}
   \frac{1}{2\pi i} \ointctrclockwise\limits_{|\omega|=\Omega} d\omega\,
   (-\omega)^{-1-2n\epsilon}
   = - \Omega^{-2n\epsilon}\,\frac{\sin 2\pi n\epsilon}{2\pi n\epsilon} \,.
   \vspace{-0.1cm}
\end{equation}

\begin{figure}
\begin{center}
\includegraphics[width=0.9\textwidth]{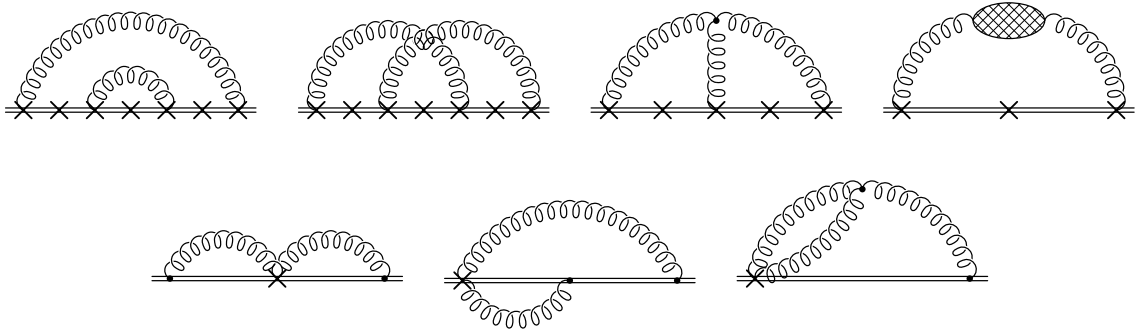} 
\end{center}
\vspace*{-0.5cm}
\caption{\label{fig:graphs}
Two-loop graphs contributing to the soft function. Double lines denote 
heavy-quark propagators, while crosses denote possible insertions of the 
operator $(\omega+in\cdot D+i0)^{-1}$.}
\end{figure}

The two-loop Feynman diagrams contributing to the matrix element in 
(\ref{eq:softdisp}) are shown in Figure~\ref{fig:graphs}. They are on-shell 
heavy-quark self-energy diagrams with an operator insertion of the 
gauge-covariant light-cone propagator. Instead of drawing a separate diagram 
for each insertion, we draw the topology for a set of diagrams and indicate 
with a cross the locations where the operator can be inserted. The loop 
integrals arising in the calculation of the soft function contain heavy-quark 
as well as light-cone propagators. The one-loop master integral is
\begin{equation}\label{eq:oneloopint}
   \int d^dk\,\frac{(-1)^{-a-b-c}}%
    {\left( k^2+i0 \right)^a \left( v\cdot k+i0 \right)^b 
     \left( n\cdot k+\omega +i0 \right)^c}
   = i\pi^\frac{d}{2}\,2^b \left( -\omega \right)^{d-2a-b-c} I_1(a,b,c) \,,
\end{equation}
where $\omega\equiv\omega+i0$, and
\begin{equation}
   I_1(a,b,c)
   = \frac{\Gamma(a+b-\frac{d}{2})\,\Gamma(2a+b+c-d)\,\Gamma(d-2a-b)}%
          {\Gamma(a)\,\Gamma(b)\,\Gamma(c)} \,.
\end{equation}
The most general two-loop loop integral we need has the form 
\begin{eqnarray}\label{eq:twoloopint}
   && \int d^dk\,d^dl\,
   \frac{(-1)^{-a_1-a_2-a_3-b_1-b_2-b_3-c_1-c_2}}%
    {\left( k^2 \right)^{a_1} \left( l^2 \right)^{a_2}
     \left[ (k-l)^2 \right]^{a_3} \left( v\cdot k \right)^{b_1}
     \left( v\cdot l \right)^{b_2} \left[ v\cdot(k+l) \right]^{b_3}
     \left( n\cdot k+\omega \right)^{c_1}
     \left( n\cdot l+\omega \right)^{c_2}} \nonumber\\
   &=& - \pi^d\,2^{b_1+b_2+b_3}
    \left( -\omega \right)^{2d-2a_1-2a_2-2a_3-b_1-b_2-b_3-c_1-c_2}
    I_2(a_1,a_2,a_3,b_1,b_2,b_3,c_1,c_2) \,,
\end{eqnarray}
where all denominators have to be supplied with a ``$+i0$'' prescription. Note 
that we do not restrict the exponents $a_1, \dots, c_2$ to be positive. Loop 
integrals with non-trivial numerators are written as linear combinations of 
integrals for which some of the indices take negative values. A third 
light-cone propagator, $[n\cdot(k-l)+\omega]^{-1}$, can be eliminated using 
partial fractioning followed by a shift of the loop momenta. We use 
integration-by-parts identities \cite{Tkachov:1981wb} to reduce the two-loop 
integrals to a minimal set of master integrals. These linear algebraic 
identities are derived by observing that integrals over total derivatives 
vanish in dimensional regularization. With 
$I_2\equiv I_2(a_1,a_2,a_3,b_1,b_2,b_3,c_1,c_2)$, the four relations obtained 
from inserting each of the differential operators $\partial_{k_\mu} k^\mu$, 
$\partial_{k_\mu} l^\mu$, $\partial_{k_\mu} n^\mu$, and 
$\partial_{k_\mu} v^\mu$ into (\ref{eq:twoloopint}) take the form
\begin{eqnarray}\label{eq:ibp}
   0 &=& \left[ d - 2a_1 - a_3 - b_1 - c_1
    - a_3\,{\bf a_1^-}\,{\bf a_3^+} + a_3\,{\bf a_2^-}\,{\bf a_3^+}
    - b_3\,{\bf b_1^-}\,{\bf b_3^+} + c_1\,{\bf c_1^+} \right ]  I_2
    \,, \nonumber\\ 
   0 &=& \left[ a_3 - a_1 - a_1\,{\bf a_1^+}\,{\bf a_2^-}
    + a_1\,{\bf a_1^+}\,{\bf a_3^-} - a_3\,{\bf a_1^-}\,{\bf a_3^+}
    + a_3\,{\bf a_2^-}\,{\bf a_3^+} \right. \nonumber\\
   &&\left. \mbox{}- b_1 \,{\bf b_1^+}\,{\bf b_2^-}
    - b_3\,{\bf b_2^-}\,{\bf b_3^+} + c_1\,{\bf c_1^+}
    - c_1\,{\bf c_1^+}\,{\bf c_2^-} \right] I_2 \,, \nonumber\\
   0 &=& \left[ a_1\,{\bf a_1^+} - a_1\,{\bf a_1^+}\,{\bf c_1^-}
    - a_3\,{\bf a_3^+}\,{\bf c_1^-} + a_3\,{\bf a_3^+}\,{\bf c_2^-}
    + b_1\,{\bf b_1^+} + b_3\,{\bf b_3^+} \right] I_2 \,, \nonumber\\
   0 &=& \left[ a_1\,{\bf a_1^+}\,{\bf b_1^-} + a_3\,{\bf a_3^+}\,{\bf b_1^-}
    - a_3\,{\bf a_3^+}\,{\bf b_2^-} - 2b_1\,{\bf b_1^+} - 2b_3\,{\bf b_3^+}
    - c_1\,{\bf c_1^+} \right] I_2 \,.
\end{eqnarray}
In these equations, the operator ${\bf a_n^+}$ (${\bf a_n^-}$) raises (lowers) 
the index $a_n$ by one unit. Four additional relations (with indices 
$1\leftrightarrow 2$ interchanged) are obtained taking derivatives with 
respect to $l_\mu$. Also, there are partial-fraction identities for integrals 
containing three different heavy-quark propagators, which follow from the 
relation ${\bf b_1^-}+{\bf b_2^-}-{\bf b_3^-}=0$. Repeated application of 
these identities can be used to set at least one of the $b_i$ exponents to 
zero.

The reduction to master integrals can be performed using computer algebra. To 
do so, one generates the equations for the index range relevant for a given 
calculation and uses Gaussian elimination to express complicated integrals in 
terms of simpler ones \cite{Laporta:2001dd}. Since the number of equations is 
very large, the order in which the equations are solved is crucial, and 
\cite{Laporta:2001dd} devised an efficient method to perform the reduction. A 
fast implementation of this algorithm is available in the form of a program 
that solves the relations (\ref{eq:ibp}) and generates a database containing the 
result for each integral \cite{Anastasiou:2004vj}. At the end of the process, 
we are left with the master integrals
\begin{eqnarray}
   M_1 &=& I_2(1,0,1,0,1,0,0,1)
    = - \frac{2^{5-2 d}\pi^3}{\sin^2(d\pi)\cos(d\pi)}\,
    \frac1{\Gamma^2(\frac{d-1}{2})} \,, \nonumber\\
   M_2 &=& I_2(1,0,1,0,1,0,1,0)
    = - \frac{\pi}{\sin(2d\pi)}\,\textstyle \Gamma(4-\frac{3d}{2})\,
    \Gamma(\frac{d}{2}-1) \,, \nonumber\\
   M_3 &=& I_2(1,0,1,1,1,0,0,1)
    = \frac{2^{2d-7}\pi^3}{\sin^2(d\pi)}\,
    \frac{\Gamma(7-2d)}{\Gamma^2(\frac{5-d}{2})} \,, \nonumber\\
   M_4 &=& I_2(1,1,0,1,1,0,1,1) = I_1(1,1,1)^2 \,.
\end{eqnarray}
The first two are evaluated by using the well-known result for the one-loop 
self-energy integral to perform the first loop integration. The result takes 
the form (\ref{eq:oneloopint}) with non-integer exponents, so that the second 
loop integration is trivial. The only master integral that needs special 
consideration is the third one. Its Feynman parameterization is
\begin{equation}
   M_3 = \Gamma(7-2d)\,\Gamma({\textstyle\frac{d}{2}}-1)^2
   \int_0^1\!dx \int_0^1\!dy\, (1-x)^{d-4} (1-y)^{3-d} y^{2 d-7}
   (1-xy)^{1-\frac{d}{2}} \,.
\end{equation}
We evaluate the parameter integral by expanding
\begin{equation}
   (1-xy)^{1-\frac{d}{2}}
   = \sum_{n=0}^\infty\,(-1)^n\,
   \frac{\Gamma(2-\frac{d}{2})}{\Gamma(n+1)\,
   \Gamma(2-n-\frac{d}{2})}\,(xy)^n
\end{equation} 
and summing up the series after the integration. 

With the integrals at hand, the evaluation of the diagrams is straightforward: 
first, each diagram is written in terms of integrals of the form 
(\ref{eq:twoloopint}), and then each integral is expressed in terms of the 
master integrals. Adding up the contributions of all graphs and expanding in 
$\epsilon=2-\frac{d}{2}$, we find
\begin{eqnarray}\label{sbare}
   s_{\rm bare}(\Omega)
   &=& 1 + \frac{Z_\alpha\alpha_s}{4\pi} \left( \frac{\Omega}{\mu}
    \right)^{-2\epsilon}\! C_F\! \left[ - \frac{2}{\epsilon^2}
    + \frac{2}{\epsilon} - \frac{\pi^2}{6}
    + \left( \frac{\pi^2}{6} + \frac23\,\zeta_3 \right) \epsilon
    - \left( \frac{\pi^4}{80} + \frac23\,\zeta_3 \right) \epsilon^2
    + {\cal O}(\epsilon^3) \right] \nonumber\\
   &&\mbox{}+ \left( \frac{Z_\alpha\alpha_s}{4\pi} \right)^2
    \left( \frac{\Omega}{\mu} \right)^{-4\epsilon} C_F\,
    \Big[ C_F K_F(\epsilon) + C_A K_A(\epsilon) + T_F n_f K_f(\epsilon) 
    \Big] + \dots \,,
\end{eqnarray}
where
\begin{equation}
   Z_\alpha = 1 - \beta_0\,\frac{\alpha_s}{4\pi\epsilon} + \dots
\end{equation}
accounts for the renormalization of the bare coupling constant, and 
\begin{equation}
   \beta_0 = \frac{11}{3}\,C_A - \frac43\,T_F n_f
\end{equation}
is the first coefficient of the $\beta$-function. Throughout, 
$\alpha_s\equiv\alpha_s(\mu)$. Note that the bare soft function is scale 
independent, since the scale dependence of $\alpha_s(\mu)$ cancels against
the explicit $\mu$ dependence. Also, since the soft function is an on-shell 
matrix element of a gauge-invariant operator it is gauge invariant. The terms 
of ${\cal O}(\epsilon)$ and ${\cal O}(\epsilon^2)$ in the one-loop result 
above are needed for the evaluation of the counter-term contributions, as 
described in the next section. The two-loop coefficients are
\begin{eqnarray}
   K_F(\epsilon)
   &=& \frac{2}{\epsilon^4} - \frac{4}{\epsilon^3}
    + \frac{2-\pi^2}{\epsilon^2} + \left( 2\pi^2 - \frac{100}{3}\,\zeta_3
    \right) \frac{1}{\epsilon} - \pi^2 - \frac{59\pi^4}{60}
    + \frac{200}{3}\,\zeta_3 + {\cal O}(\epsilon) \,, \nonumber\\
   K_A(\epsilon)
   &=& - \frac{11}{6\epsilon^3}
    + \left( - \frac{1}{18} + \frac{\pi^2}{6} \right) \frac{1}{\epsilon^2}
    + \left( - \frac{55}{27} - \frac{23\pi^2}{36} + 9\zeta_3 \right)
    \frac{1}{\epsilon} \nonumber\\
   &&\mbox{}- \frac{326}{81} - \frac{361\pi^2}{108} + \frac{67\pi^4}{180}
    - \frac{85}{9}\,\zeta_3 + {\cal O}(\epsilon) \,, \nonumber\\
   K_f(\epsilon)
   &=& \frac{2}{3\epsilon^3} - \frac{2}{9\epsilon^2}
    + \left( - \frac{4}{27} + \frac{\pi^2}{9} \right) \frac{1}{\epsilon}
    - \frac{8}{81} - \frac{\pi^2}{27} - \frac{28}{9}\,\zeta_3
    + {\cal O}(\epsilon) \,.
\end{eqnarray}

\section{Renormalization of the soft function\label{sec:renormalization}}

As usual, we define an operator renormalization factor $Z$ via
\begin{equation}
   S(\omega,\mu) = \int d\omega'\,Z(\omega,\omega',\mu)\,
   S_{\rm bare}(\omega') \,,
\end{equation}
where $Z$ absorbs the UV divergences of the bare soft function, such that the
renormalized soft function is finite in the limit $\epsilon\to 0$. Here and 
below, all integrals over variable $\omega$, $\omega'$ etc.\ run from 0 to 
$\infty$. The convolution integral can be understood as a generalization of 
the matrix formula $O_i=Z_{ij}\,O_j^{\rm bare}$, and so the usual relation 
between the $Z$ factor and the anomalous dimension holds. It follows that
\begin{equation}\label{gamdef}
   \gamma(\omega,\omega',\mu) = - \int d\omega''\,
   \frac{dZ(\omega,\omega'',\mu)}{d\ln\mu}\,
   Z^{-1}(\omega'',\omega',\mu) \,.
\end{equation}
Below, we will write convolution integrals of this form using the short-hand
notation $\gamma=-(dZ/d\ln\mu)\otimes Z^{-1}$. In the $\overline{\rm MS}$ 
scheme, we have
\begin{equation}
   Z(\omega,\omega',\mu) = \delta(\omega-\omega')
   + \sum_{k=1}^\infty\,\frac{1}{\epsilon^k}\,Z^{(k)}(\omega,\omega',\mu) \,.
\end{equation}

The $Z$ factors depend on $\mu$ both implicitly via $\alpha_s(\mu)$, and 
explicitly via so-called star distributions related to the presence of Sudakov 
double logarithms. The latter dependence is a new feature, which leads to 
modifications of the standard relations derived, e.g., in 
\cite{Floratos:1977au}. Indeed, from the definition (\ref{gamdef}) of the 
anomalous dimension it follows that
\begin{equation}
   \gamma + \gamma\otimes \sum_{k=1}^\infty\,\frac{Z^{(k)}}{\epsilon^k}
   = - \sum_{k=1}^\infty\,\frac{1}{\epsilon^k}
   \left[ \frac{\partial Z^{(k)}}{\partial\alpha_s}\,
   \frac{d\alpha_s}{d\ln\mu} + \frac{\partial Z^{(k)}}{\partial\ln\mu}
   \right] .
\end{equation}
Here $d\alpha_s/d\ln\mu=\beta(\alpha_s)-2\epsilon\,\alpha_s$ is the 
generalized $\beta$-function in the regularized theory, and $\beta(\alpha_s)$ 
is the ordinary $\beta$-function. Both $\gamma$ and $\beta$ are independent of 
$\epsilon$. Comparing coefficients of $1/\epsilon^k$, we then find
\begin{equation}\label{gamma}
   \gamma = 2\alpha_s\,\frac{\partial Z^{(1)}}{\partial\alpha_s} \,,
\end{equation}
and
\begin{equation}\label{rela}
   2\alpha_s\,\frac{\partial Z^{(n+1)}}{\partial\alpha_s}
   = 2\alpha_s\,\frac{\partial Z^{(1)}}{\partial\alpha_s}\otimes Z^{(n)}
   + \beta(\alpha_s)\,\frac{\partial Z^{(n)}}{\partial\alpha_s}
   + \frac{\partial Z^{(n)}}{\partial\ln\mu} \,;\quad n\ge 1 \,.
\end{equation}
While (\ref{gamma}) is a familiar result \cite{Floratos:1977au}, the relations 
(\ref{rela}) contain the additional $\partial Z^{(n)}/\partial\ln\mu$ piece, 
which is usually not present.

We now use the fact that, to all orders in perturbation theory, the anomalous 
dimension of the shape function is given by 
\cite{Neubert:2004dd,Grozin:1994ni}
\begin{equation}
   \gamma(\omega,\omega',\mu)
   = -2\Gamma_{\rm cusp}(\alpha_s) \left( \frac{1}{\omega-\omega'}
   \right)_{\!*}^{\![\mu]} + 2\gamma(\alpha_s)\,\delta(\omega-\omega') \,.
\end{equation}
Here and below we encounter star distributions defined as 
\cite{DeFazio:1999sv}
\begin{eqnarray}
   \int_{\le 0}^\Omega\!d\omega\,f(\omega) 
   \left( \frac{1}{\omega} \right)_{\!*}^{\![\mu]}
   &=& \int_0^\Omega\!d\omega\,\frac{f(\omega)-f(0)}{\omega}
    + f(0)\,\ln\frac{\Omega}{\mu} \,, \nonumber\\
   \int_{\le 0}^\Omega\!d\omega\,f(\omega) 
   \left( \frac{\ln\frac{\omega}{\mu}}{\omega} \right)_{\!*}^{\![\mu]}
   &=& \int_0^\Omega\!d\omega\,\frac{f(\omega)-f(0)}{\omega}\,
    \ln\frac{\omega}{\mu}
    + \frac{f(0)}{2}\,\ln^2\frac{\Omega}{\mu} \,,
\end{eqnarray}
where $f(\omega)$ is a smooth test function. It follows from (\ref{gamma}) 
that
\begin{equation}
   Z^{(1)}(\omega,\omega',\mu)
   = -2 Z_{\rm cusp}^{(1)}(\alpha_s) \left( \frac{1}{\omega-\omega'}
   \right)_{\!*}^{\![\mu]} + 2Z_\gamma^{(1)}(\alpha_s)\,
   \delta(\omega-\omega') \,,
\end{equation}
where
\begin{eqnarray}
   \Gamma_{\rm cusp}(\alpha_s)
   &=& 2\alpha_s\,\frac{\partial Z_{\rm cusp}^{(1)}}{\partial\alpha_s}
    = \sum_{n=0}^\infty\,\Gamma_n \left( \frac{\alpha_s}{4\pi} \right)^{n+1}
    , \nonumber\\
   \gamma(\alpha_s)
   &=& 2\alpha_s\,\frac{\partial Z_\gamma^{(1)}}{\partial\alpha_s}
    = \sum_{n=0}^\infty\,\gamma_n \left( \frac{\alpha_s}{4\pi} \right)^{n+1} .
\end{eqnarray}
$\Gamma_{\rm cusp}$ is the cusp anomalous dimension already mentioned in 
connection with (\ref{etadef}), whose two-loop expression has been derived in 
\cite{Korchemsky:1987wg}. The other anomalous dimension, $\gamma$, has been 
calculated at two-loop order in \cite{Korchemsky:1992xv,Gardi:2005yi}. The 
relevant expansion coefficients are
\begin{eqnarray}\label{Gamman}
   \Gamma_0 &=& 4 C_F \,, \hspace{1.1cm}
    \Gamma_1 = C_F \left[ \left( \frac{268}{9} - \frac{4\pi^2}{3} \right)
    C_A - \frac{80}{9}\,T_F n_f \right] , \nonumber\\
   \gamma_0 &=& - 2 C_F  \,, \qquad
    \gamma_1 = C_F \left[ \left( \frac{110}{27} + \frac{\pi^2}{18}
    - 18\zeta_3 \right) C_A
    + \left( \frac{8}{27} + \frac{2\pi^2}{9} \right) T_F n_f \right] .
\end{eqnarray}

The relations (\ref{rela}) now allow us to express the coefficients $Z^{(k)}$ 
in terms of the expansion coefficients of $\beta$, $\Gamma_{\rm cusp}$, and 
$\gamma$. To derive these results, we need the following identities for 
star distributions (the first of which is somewhat laborious to derive)
\begin{eqnarray}
   \left( \frac{1}{\omega-\omega''} \right)_{\!*}^{\![\mu]}
   \otimes \left( \frac{1}{\omega''-\omega'} \right)_{\!*}^{\![\mu]}
   &=& 2 \left( \frac{\ln\frac{\omega-\omega'}{\mu}}{\omega-\omega'}
    \right)_{\!*}^{\![\mu]} - \frac{\pi^2}{6}\,\delta(\omega-\omega') \,,
    \nonumber\\
   \frac{d}{d\ln\mu} \left( \frac{1}{\omega-\omega'} \right)_{\!*}^{\![\mu]}
   &=& - \delta(\omega-\omega') \,, \nonumber\\
   \frac{d}{d\ln\mu} \left(
    \frac{\ln\frac{\omega-\omega'}{\mu}}{\omega-\omega'}
    \right)_{\!*}^{\![\mu]}
   &=& -  \left( \frac{1}{\omega-\omega'} \right)_{\!*}^{\![\mu]} \,.
\end{eqnarray}
Denoting by $Z_{[n]}$ the coefficient of $(\alpha_s/4\pi)^n$ in 
$Z(\omega,\omega',\mu)$, we obtain after some algebra
\begin{eqnarray}
   Z_{[0]} &=& \delta(\omega-\omega') \,, \nonumber\\
   Z_{[1]} &=& \delta(\omega-\omega')
    \left( \frac{\Gamma_0}{2\epsilon^2} + \frac{\gamma_0}{\epsilon} \right)
    - \frac{\Gamma_0}{\epsilon}
    \left( \frac{1}{\omega-\omega'} \right)_{\!*}^{\![\mu]} \,, \nonumber\\
   Z_{[2]} &=& \delta(\omega-\omega')
    \left[ \frac{\Gamma_0^2}{8\epsilon^4}
    + \frac{\Gamma_0(\gamma_0 - \frac34\beta_0)}{2\epsilon^3}
    + \left( \frac{\gamma_0(\gamma_0-\beta_0)}{2} + \frac{\Gamma_1}{8}
    - \frac{\pi^2}{12}\,\Gamma_0^2 \right) \frac{1}{\epsilon^2}
    + \frac{\gamma_1}{2\epsilon} \right] \nonumber\\
   &&\mbox{}- \left( \frac{1}{\omega-\omega'} \right)_{\!*}^{\![\mu]}
    \left[ \frac{\Gamma_0^2}{2\epsilon^3}
    + \frac{\Gamma_0(\gamma_0 - \frac12\beta_0)}{\epsilon^2}
    + \frac{\Gamma_1}{2\epsilon} \right]
    + \frac{\Gamma_0^2}{\epsilon^2}
    \left( \frac{\ln\frac{\omega-\omega'}{\mu}}{\omega-\omega'}
    \right)_{\!*}^{\![\mu]} \,.
\end{eqnarray}
This is the complete two-loop result for the renormalization factor of the 
$B$-meson shape function. 

According to (\ref{jdefsdef}), the soft function is defined as the integral 
over the renormalized (parton-model) shape function. Using the fact that 
$Z(\omega,\omega',\mu)$ only depends on the difference $(\omega-\omega')$, 
we find that
\begin{equation}
   s\Big( \frac{\Omega}{\mu},\mu \Big)
   = \int_0^\Omega\!d\omega\,Z(\Omega,\omega,\mu)\,s_{\rm bare}(\omega) \,,
\end{equation}
where $s_{\rm bare}(\Omega)$ is defined as the integral over the bare shape
function and is scale independent. Expanding this relation in perturbation 
theory, we obtain
\begin{eqnarray}
   s_{[0]} &=& s_{[0]}^{\rm bare} \,, \nonumber\\
   s_{[1]} &=& Z_{[0]}\otimes s_{[1]}^{\rm bare} 
    + Z_{[1]}\otimes s_{[0]}^{\rm bare} \,, \nonumber\\
   s_{[2]} &=& Z_{[0]}\otimes s_{[2]}^{\rm bare}
    + Z_{[1]}\otimes s_{[1]}^{\rm bare} + Z_{[2]}\otimes s_{[0]}^{\rm bare}
    \,,
\end{eqnarray}
with $s_{[0]}^{\rm bare}=1$. The first term on the right-hand side in each 
line corresponds to the result obtained from the loop diagrams, given in 
(\ref{sbare}). The remaining terms correspond to operator counter-terms. 
Explicitly, we obtain for the counter-term contributions
\begin{eqnarray}\label{opsCT}
   s_{[1]}^{\rm C.T.}
   &=& \frac{\Gamma_0}{2\epsilon^2} + \frac{\gamma_0}{\epsilon}
    - \frac{\Gamma_0}{\epsilon}\,\ln\frac{\Omega}{\mu} \,, \nonumber\\
   s_{[2]}^{\rm C.T.}
   &=& \left[ \frac{\Gamma_0}{2\epsilon^2}
    + \frac{\gamma_0}{\epsilon} - \frac{\Gamma_0}{\epsilon} \left(
    \ln\frac{\Omega}{\mu} - H_{-2\epsilon} \right) \right]
    s_{[1]}^{\rm bare}(\Omega) \nonumber\\
   &&\mbox{}+ \frac{\Gamma_0^2}{8\epsilon^4}
    + \frac{\Gamma_0(\gamma_0 - \frac34\beta_0)}{2\epsilon^3}
    + \left( \frac{\gamma_0(\gamma_0-\beta_0)}{2} + \frac{\Gamma_1}{8}
    - \frac{\pi^2}{12}\,\Gamma_0^2 \right) \frac{1}{\epsilon^2}
    + \frac{\gamma_1}{2\epsilon} \nonumber\\
   &&\mbox{}- \left[ \frac{\Gamma_0^2}{2\epsilon^3}
    + \frac{\Gamma_0(\gamma_0 - \frac12\beta_0)}{\epsilon^2}
    + \frac{\Gamma_1}{2\epsilon} \right] \ln\frac{\Omega}{\mu}
    + \frac{\Gamma_0^2}{2\epsilon^2}\,\ln^2\frac{\Omega}{\mu} \,,
\end{eqnarray}
where $H_{-2\epsilon}$ is the harmonic number, which results form the
integral
\begin{equation}
   \int_0^1\!dx\,\frac{1-x^{-2\epsilon}}{1-x} = H_{-2\epsilon} \,.
\end{equation}

The counter-term contributions can be evaluated using the results for the 
bare one-loop soft function from (\ref{sbare}) and the expressions for the 
anomalous-dimension coefficients given in (\ref{Gamman}). When adding these 
contributions to the result (\ref{sbare}) for the bare soft function we find 
that all $1/\epsilon^n$ pole terms cancel, so that the limit $\epsilon\to 0$ 
can now be taken. In \cite{Neubert:2005nt} the logarithmic terms in the 
renormalized soft function have been determined by solving the 
renormalization-group equation for the function $s$. At two-loop order, it was 
found that
\begin{eqnarray}\label{jands}
   s(L,\mu) &=& 1 + \frac{\alpha_s(\mu)}{4\pi} \left[
    c_0^{(1)} + 2\gamma_0 L - \Gamma_0 L^2 \right] \nonumber\\
   &&\mbox{}+ \left( \frac{\alpha_s(\mu)}{4\pi} \right)^2 \Bigg[
    c_0^{(2)} + \left( 2c_0^{(1)} (\gamma_0-\beta_0) + 2\gamma_1
    + \frac{2\pi^2}{3}\,\Gamma_0\gamma_0 + 4\zeta_3\Gamma_0^2 \right) L \\
   &&\mbox{}+ \left( 2\gamma_0(\gamma_0-\beta_0) - c_0^{(1)}\Gamma_0
    - \Gamma_1 - \frac{\pi^2}{3}\,\Gamma_0^2 \right) L^2 
    + \left( \frac23\,\beta_0 - 2\gamma_0 \right) \Gamma_0 L^3
    + \frac{\Gamma_0^2}{2}\,L^4 \Bigg] \,, \nonumber
\end{eqnarray}
where the non-logarithmic one-loop coefficient reads 
\cite{Bosch:2004th,Bauer:2003pi}
\begin{equation}
   c_0^{(1)} = - \frac{\pi^2}{6}\,C_F \,.
\end{equation}
Our results for the logarithmic terms agree with (\ref{jands}) and thus 
confirm the existing results for the two-loop anomalous dimensions $\Gamma_1$ 
\cite{Korchemsky:1987wg} and $\gamma_1$ \cite{Korchemsky:1992xv,Gardi:2005yi}. 
In addition, our calculation gives for the non-logarithmic piece at two-loop 
order the expression
\begin{eqnarray}
   c_0^{(2)} &=& C_F^2 \left( - \frac{4\pi^2}{3} - \frac{3\pi^4}{40}
    + 32\zeta_3 \right) 
    + C_F C_A \left( - \frac{326}{81} - \frac{427\pi^2}{108}
    + \frac{67\pi^4}{180} - \frac{107}{9}\,\zeta_3 \right) \nonumber\\
   &&\mbox{}+ C_F T_F n_f \left( - \frac{8}{81} + \frac{5\pi^2}{27}
    - \frac{20}{9}\,\zeta_3 \right) .
\end{eqnarray}
This is the main result of the present work. 

It is interesting to compare the exact answer for the coefficient $c_0^{(2)}$ 
with the approximation obtained by keeping only the terms of order 
$\beta_0\alpha_s^2$, which could be derived without any of the elaborate 
technology developed in the present paper. In the absence of exact two-loop 
results, it is sometimes argued that the $\beta_0\alpha_s^2$ terms constitute 
the dominant part of the complete two-loop correction. In the present case, we 
obtain for $N_c=3$ colors (note that $\beta_0=9$ for $n_f=3$ light flavors)
\begin{equation}
   c_0^{(2)}\approx 8.481\cdot\frac{\beta_0}{9} - 62.682
   \approx -54.201 \,.
\end{equation}
Keeping only the $\beta_0\alpha_s^2$ term would give $8.481$, which has the 
wrong sign and is off by almost an order of magnitude. This illustrates the 
importance of performing exact two-loop calculations.

\section{Discussion and summary}
\label{sec:concl}

\begin{figure}
\begin{center}
\includegraphics[width=0.9\textwidth]{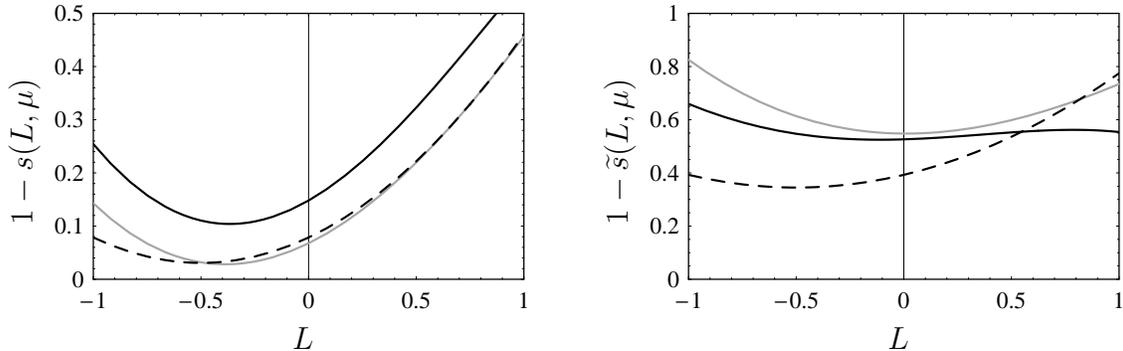}
\end{center}
\vspace{-0.5cm}
\caption{ \label{fig:softfunction}
One- and two-loop corrections to the soft functions $s(L,\mu)$ and 
$\widetilde s(L,\mu)$ evaluated at $\alpha_s(\mu)=0.45$. The dashed lines show 
the one-loop results results, while the solid lines give the complete two-loop 
results derived in the present work. The gray lines are obtained if only the 
$\beta_0\alpha_s^2$ terms are kept in the two-loop contributions.}
\end{figure}

Having completed the two-loop calculation of the soft function, we now briefly 
discuss the impact of our results for the prediction of the partial inclusive 
$\bar B\to X_s\gamma$ branching ratio. We begin by displaying the final 
expressions for the functions $s(L,\mu)$ and $\widetilde s(L,\mu)$ obtained 
from (\ref{jands}) and (\ref{replace}) for the case of $N_c=3$ colors and 
$n_f=3$ light quark flavors. We obtain
\begin{eqnarray}
   s(L,\mu)
   &\approx& 1 + \left( -0.175 - 0.424 L - 0.424 L^2\right) \alpha_s(\mu)
    \nonumber\\
   &&\mbox{}+ \left( -0.343 - 0.201 L - 0.433 L^2 + 0.383 L^3 + 
   0.090 L^4  \right) \alpha_s^2(\mu) + \dots \,, \nonumber\\
   \widetilde s(L,\mu)
   &\approx& 1 + \left( -0.873 - 0.424 L - 0.424 L^2 \right) \alpha_s(\mu)
    \nonumber\\
   &&\mbox{}+ \left( -0.660 + 0.821 L + 0.456 L^2 + 0.383 L^3 + 
   0.090 L^4 \right) \alpha_s^2(\mu) + \dots \,.
\end{eqnarray}
The two-loop corrections are quite significant, especially in the case of 
$\widetilde s(L,\mu)$. Figure~\ref{fig:softfunction} shows the dependence of 
the soft functions on $L=\ln(\Omega/\mu)$ at the fixed renormalization scale 
$\mu$ chosen such that $\alpha_s(\mu)=0.45$, corresponding to a 
renormalization point $\mu\approx 1.1$\,GeV as appropriate for the 
photon-energy cuts used in current measurements of the $\bar B\to X_s\gamma$ 
decay. In addition to the one- and two-loop results, we display the results 
obtained if only terms of order $\beta_0\alpha_s^2$ are kept in the two-loop 
coefficients. The figure shows that the two-loop effects calculated in this 
paper can have an impact on the soft functions at the 10--20\% level, and that 
keeping only the $\beta_0\alpha_s^2$ terms does, in general, not provide an 
accurate description of the two-loop effects. We stress, however, that the 
large perturbative corrections seen in the figure do not translate in 
similarly large corrections to the $\bar B\to X_s\gamma$ decay rate. The size 
of the corrections is strongly reduced once the pole mass in (\ref{GOPE}) is 
eliminated in favor of a low-scale subtracted $b$-quark mass, such as the 
shape-function mass \cite{Bosch:2004th,Neubert:2004sp}. At one-loop order this 
was demonstrated in \cite{Neubert:2004dd}, and we expect similar cancelations 
to persist in higher orders. Note also that the soft functions by themselves 
are not renormalization-group invariant, so it is meaningless to study their 
dependence on the scale $\mu$ for fixed $\Omega$. In physical results such as 
the expression for the $\bar B\to X_s\gamma$ decay rate in (\ref{GOPE}), the 
scale dependence of the soft function cancels against the $\mu_0$ dependence 
of the objects $U_2$, $(\Delta/\mu_0)^\eta$, and $\eta$ in (\ref{etadef}). A 
detailed analysis of the phenomenological impact of NNLO corrections on the 
$\bar B\to X_s\gamma$ decay rate will be given elsewhere.

In summary, we have calculated the two-loop expression for soft function 
$s(L,\mu)$, which is defined in terms of an integral over the $B$-meson shape 
function in the parton model. This quantity is a necessary ingredient for the 
NNLO evaluation of the $\bar B\to X_s\gamma$ decay rate with a cut on the 
photon energy. For a sufficiently low cut energy, the partial inclusive decay 
rate can be calculated in a multi-scale operator-product expansion, in which 
the soft function arises in the final expansion step, when a current-current 
correlator in soft-collinear effective theory is matched onto bilocal 
heavy-quark operators in heavy-quark effective theory. If the cut on the 
photon energy is so severe that the contribution from the soft region cannot 
be evaluated perturbatively, the soft part should be subtracted from the 
partonic result for the cut rate, before it is convoluted with the 
renormalized shape function. Even in that case our result is a necessary 
component in the consistent calculation of the $\bar B\to X_s\gamma$ photon
spectrum at NNLO. Moreover, since the soft function is universal to all 
inclusive heavy-to-light decays in the end-point region, our results are also 
relevant for semi-leptonic $\bar B\to X_u l^-\bar\nu$ decay spectra, if one 
attempts to extend the analysis of \cite{Bosch:2004th,Lange:2005yw} to NNLO.

\subsection*{Acknowledgments}

We are grateful to the Institute for Theoretical Physics at the University of 
Heidelberg, Germany, for its hospitality during the completion of much of this 
work. This research of T.B.\ was supported by the Department of Energy under 
Grant DE-AC02-76CH03000. The research of M.N.\ was supported in part by a 
Research Award of the Alexander von Humboldt Foundation, and by the National 
Science Foundation under Grant PHY-0355005. Fermilab is operated by 
Universities Research Association Inc., under contract with the U.S.\ 
Department of Energy.

\end{document}